\documentclass[aip,jap,reprint,amsmath,amssymb]{revtex4-2}

\usepackage{graphicx}
\usepackage{bm}
\usepackage{booktabs}
\usepackage{array}
\usepackage{enumitem}

\newcommand{\Psucc}{P_{\rm success}}
\newcommand{\Ceff}{C_{\rm effective}}
\newcommand{\Reall}{\mathrm{Re}}

\begin{document}
	
	\title{Attractor-Basin-Limited Fidelity in Reproducible Multistate Vortex Memory}
	
	\author{M. Rabiu, S. S. Abukari, M. Amekpewu}
	\affiliation{Department of Physics, Faculty of Physical Sciences, University for Development Studies, Tamale, Ghana}
	\email{rmusah@uds.edu.gh}
	
	\date{\today}
	
	\begin{abstract}
	Multilevel non-volatile memory technologies face recurring trade-offs among information density, endurance, retention, and switching energy. We investigate an alternative state variable based on the discrete vortex configuration of self-organized vortices in a boundary-driven electron fluid. A dissipative point-vortex model derived from magnetohydrodynamic dynamics yields six reproducible vortex codewords, corresponding to 2.585 bits per cell, whose write fidelity, noise sensitivity, 100-cycle endurance, and effective information capacity are quantified. The ordering of their finite-amplitude basin radii $r_{50}$ differs from that predicted by both fixed-circulation and fully coupled linear-stability spectra. A nonlinear saddle-point construction based on the reduced dynamics likewise does not recover the measured basin ordering. The discrepancy is associated with escape pathways involving coupled position–circulation dynamics that are absent from the fixed-circulation description. These results show that local stability alone does not determine the finite perturbation tolerance of the vortex states considered here. The proposed memory requires active hold power, and passive retention remains to be established experimentally.
	\end{abstract}
	
	\maketitle
	
	\section{Introduction}
	\label{sec:intro}
	
	Non-volatile memory technology faces a structural engineering constraint: emerging replacements—phase-change memory (PCM), resistive RAM (RRAM), and spin-transfer-torque MRAM—encode information in a physical variable that supports only a few distinguishable levels, and pushing the level count higher degrades endurance, retention, or switching energy \citep{ye2021overview}. A route to high-endurance multilevel storage is therefore to use a state variable whose distinguishable states do not rely solely on finely spaced energy barriers.
	
	Vortices are attractive candidates because their circulation and spatial organization provide discrete configurational degrees of freedom that can support robust multistability \citep{golod2015single,xiong2022ferroelectric,goto2021twin}. In two-dimensional electron fluids governed by magnetohydrodynamic (MHD) boundary conditions, vortices can be generated and controlled by electromagnetic driving \citep{bozkaya2011numerical}, and high-mobility two-dimensional systems—graphene in particular—now support hydrodynamic electron flow \citep{bandurin2016negative,krishnakumar2017superballistic}. This suggests a route to a vortex memory whose state space is a self-organized attractor landscape.
	
	A memory state is considered operationally useful when it is dynamically reachable, reproducible, distinguishable, and robust to finite perturbations. The analysis focuses on the attractors that can be written repeatedly across independent initializations and on the geometry of their basins of attraction. For each retained state we quantify write/read confusion, sensitivity to distinct perturbation channels, finite-amplitude basin robustness, local stability, retention, cycling behavior, and effective information capacity.
	
	The calculations use a reduced dissipative point-vortex representation of the boundary-driven MHD system. The model resolves multistable vortex dynamics and permits systematic sampling of write protocols and perturbations, but it is not a materials-calibrated device simulation. Accordingly, the reported noise thresholds, basin radii, cycle counts, and control-work values characterize the reduced model. Continuum-PDE validation, materials-specific parameter calibration, physical switching-energy measurements, and experimental device operation remain outside the present scope.
	
	\section{Physical model and simulation results}
	\label{sec:results}
	
	\subsection{Model and attractor discovery}
	\label{sec:model}
	
\begin{figure*}[t]
		\centering
		\includegraphics[width=\textwidth]{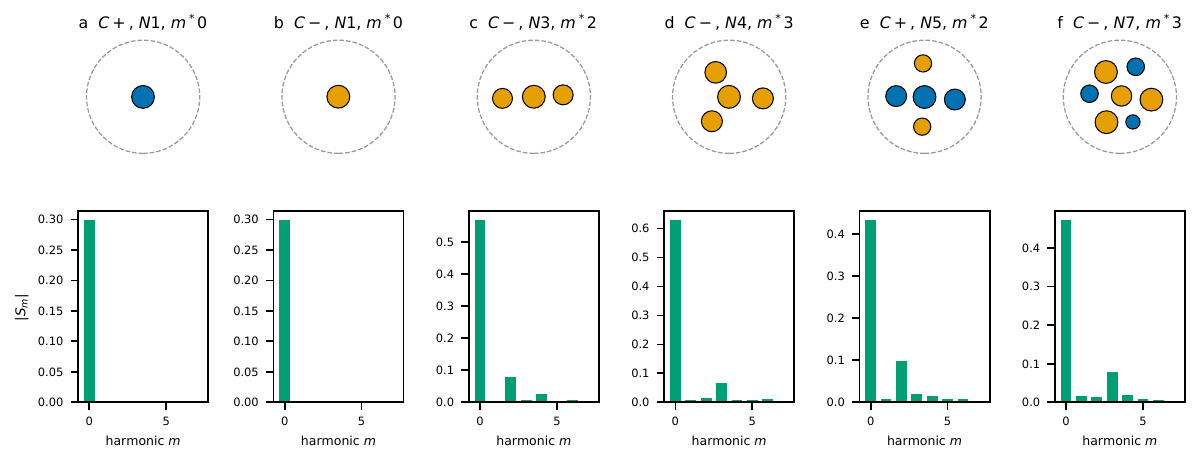}
		\caption{Device schematic and held-state snapshots (top) and far-field boundary-harmonic spectra (bottom) of the six confirmed codewords. \textbf{a}, Disk with rim current $I(t)$ and $P$ magnetization wells. \textbf{b–g}, Cluster centroids and far-field multipole spectrum $|S_m|$ for codewords S0–S5.}
		\label{fig:fig1}
	\end{figure*}

	The device is a boundary-driven point-vortex system: a disk of radius $R=1$ carrying an azimuthal rim current $I(t)$ and $P$ Gaussian magnetization wells of signed amplitude $b_w s_p$ ($s_p\in\{\pm1\}$) at radius $0.55R$ (Fig.~\ref{fig:fig1}a). Circulation is injected as a jittered 16-vortex rim ring; satellite vortices are seeded at the $P$ well locations during the split phase; and the dissipative point-vortex equations of motion (Biot–Savart plus disk Milne–Thomson images, a rim-current solid-body rotation, local Hartmann-plus-viscous damping, and magnetic-pinning guiding-center drift) are integrated with RK4 (Appendix~\ref{app:methods}). Every held state is classified by the triple $\sigma=(\mathrm{sign}(C), N, m^*)$, where $C=\Gamma_{\rm tot}$ is the net circulation, $N$ is the number of spatially coherent clustered vortex cores, and $m^*$ is the dominant non-axisymmetric harmonic of the far-field circulation-multipole spectrum (Appendix~\ref{app:methods}); for $N\le1$ we set $m^*\equiv0$.
	
	We swept $P\in\{0,2,3,4,5,6\}$, well amplitude $b_w\in\{1,2,3\}$, write current $|I_{\rm write}|\in\{0.8,1.2,1.6\}$ (both polarities), and well-sign pattern (alternating/all-same): 186 independent write(0.25)–split(0.15)–hold(0.6) protocols at seed 0. This produced 49 distinct signatures $\sigma$. We selected the 24 most frequent signatures and re-ran each at four additional independent seeds. A candidate is accepted if the identical signature is recovered in at least four of five seeds. Seven candidates passed; one is a near-mirror duplicate, so we retain six structurally diverse attractors ($N=1,1,3,4,5,7$; Table~\ref{tab:codewords}). Figure~\ref{fig:fig1} shows the device schematic and the held-state snapshots and spectral fingerprints of all six confirmed codewords.
	
	\begin{table*}[t]
		\centering
		\small
		\caption{Confirmed codewords. $\sigma=(\mathrm{sign}\,C,\,N,\,m^*)$; hits/seeds is the reproducibility count out of 5; $\Gamma,\Omega$ are per-component medians.}
		\label{tab:codewords}
		\begin{tabular}{lccccccccc}
			\toprule
			label & $\sigma$ & $P$ & $b_w$ & $I_{\rm write}$ & pattern & hits/5 & $\Gamma$ \quad $\Omega$ & Write success & Misclassification  \\
			\midrule
			S0 & $(+,1,0)$ & 0 (off) & --  & $+0.8$ & alternating & 5/5 & $\phantom{-}0.298$ \quad $0.709$ & 74\% & 26\% \\
			S1 & $(-,1,0)$ & 0 (off) & --  & $-0.8$ & alternating & 5/5 & $-0.298$ \quad $0.709$ & 80\% & 20\% \\
			S2 & $(-,3,2)$ & 2 & 1.0 & $-1.6$ & all-same     & 5/5 & $-0.709$ \quad $37.47$ & 70\% & 30\% \\
			S3 & $(-,4,3)$ & 3 & 1.0 & $-1.2$ & all-same  & 5/5 & $-0.628$ \quad $5.619$ & 100\% & 0\% \\
			S4 & $(+,5,2)$ & 4 & 1.0 & $+1.6$ & alternating  & 5/5 & $\phantom{-}0.573$ \quad $51.60$ & 72\% & 28\% \\
			S5 & $(-,7,3)$ & 6 & 1.0 & $-1.6$ & alternating  & 5/5 & $-0.625$ \quad $48.32$ & 82\% & 18\% \\
			\bottomrule
		\end{tabular}
	\end{table*}
	
	\subsection{Write/read performance and robustness}
	\label{sec:performance}
	
\begin{figure*}[t]
		\centering
		\includegraphics[width=\textwidth]{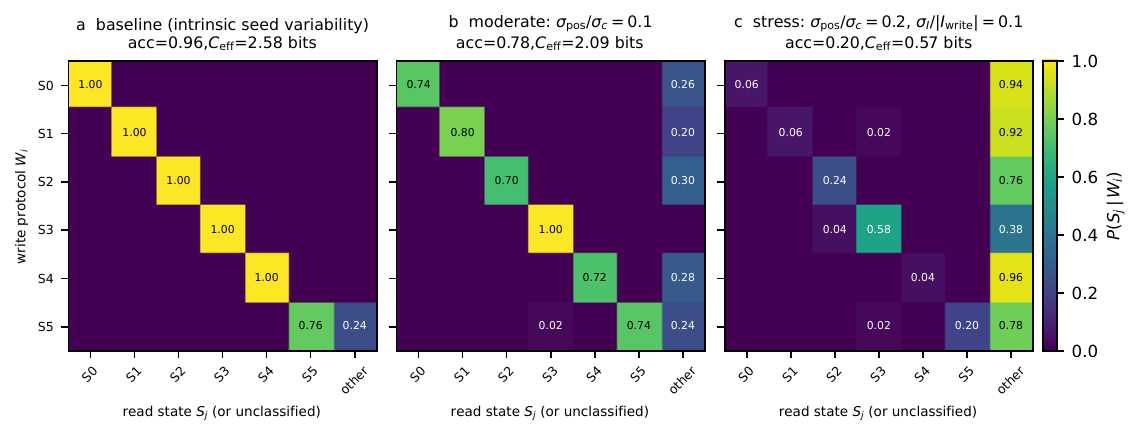}
		\caption{Write/read confusion matrices, $P(S_j\,|\,W_i)$, $n=50$ trials per row, at three operating points: \textbf{a}, baseline; \textbf{b}, moderate positional noise; \textbf{c}, combined stress. ``other'' = uncataloged signature. Off-diagonal confusion among cataloged codewords is zero at baseline and moderate noise, and small (1.3\%) at stress.}
		\label{fig:fig2}
	\end{figure*}

	We evaluate the write/read channel through a $6\times6$ confusion matrix. For each write protocol $W_i$ we ran $n=50$ independent-seed trials and classified the resulting held signature into one of the six cataloged codewords or an ``other/uncataloged'' bin, at three operating points: (a) baseline; (b) moderate positional noise ($\sigma_{\rm pos}/\sigma_c=0.1$); and (c) combined stress ($\sigma_{\rm pos}/\sigma_c=0.2$ and $\sigma_I/|I_{\rm write}|=0.1$ simultaneously). Figure~\ref{fig:fig2} shows the matrices.
	
	At baseline, every write protocol reproduces its target codeword with $P(S_i\,|\,W_i)=1.00^{+0.00}_{-0.07}$ (95\% Wilson interval). At moderate noise, diagonal accuracy falls unevenly, from $1.00$ (S3) down to $0.70$ (S2). At combined stress, every codeword's diagonal entry collapses, from $0.58$ (S3) down to $0.04$ (S4) (mean $0.18$). Off-diagonal entries among the six cataloged codewords are zero at baseline and moderate noise; at stress, only $4/300$ trials show inter-codeword confusion. Most misclassified trials land in the ``other'' column—an uncataloged attractor, not a sibling codeword.
	
\begin{figure*}[t]
		\centering
		\includegraphics[width=\textwidth]{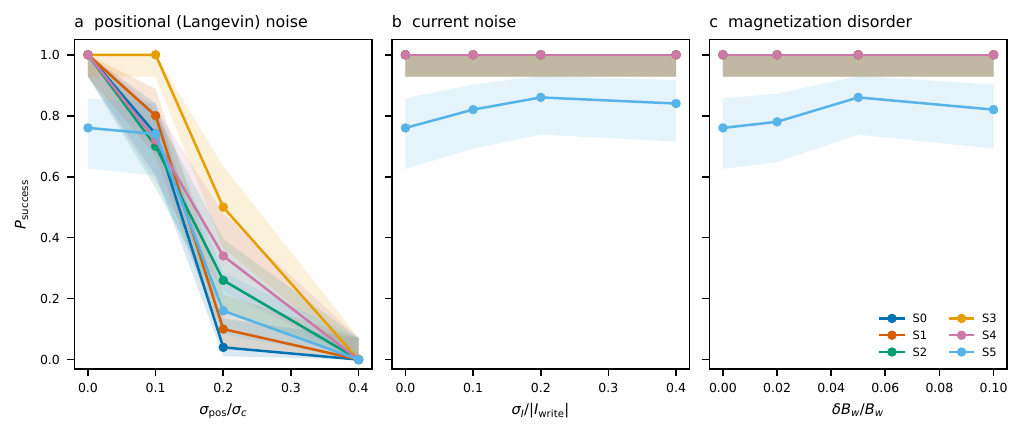}
		\caption{Write-success probability $\Psucc$ ($n=50$ trials/point, shaded bands = 95\% Wilson score CI) for all six codewords against three noise channels: \textbf{a}, positional; \textbf{b}, current; \textbf{c}, magnetization disorder. Positional noise is the only degrading channel.}
		\label{fig:fig3}
	\end{figure*}

	Figure~\ref{fig:fig3} reports $\Psucc$ for all six codewords against three noise channels: positional Langevin kicks, multiplicative write-current noise, and magnetization-well amplitude disorder. Positional noise is the dominant fragility channel; current noise and magnetization disorder leave $\Psucc=1.00$ for all six codewords at every tested amplitude.
	
\begin{figure*}[t]
		\centering
		\includegraphics[width=\textwidth]{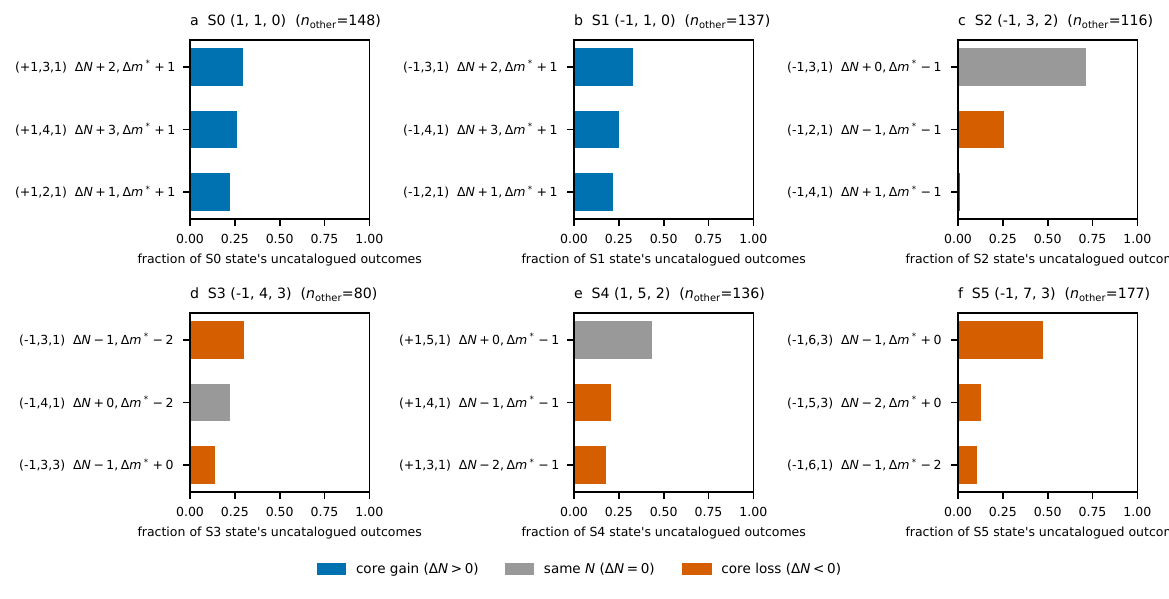}
		\caption{Hidden-attractor transition structure. For each cataloged state \textbf{a–f}, the top three recurrent uncataloged destinations, colored by change in vortex number $\Delta N$: blue = core gain, grey = same, vermillion = core loss. Most destinations have relaxed to $m^*=1$.}
		\label{fig:fig9}
	\end{figure*}

	The ``other'' category is not diffuse: among $783$ misclassified trials, only $45$ distinct signatures appear, and the three most recurrent account for $40.7\%$ of uncataloged outcomes. The dominant transition rule is a collapse toward the lowest nonzero harmonic ($m^*=1$), with $\Delta m^*\le0$ for all but two isolated exceptions. Figure~\ref{fig:fig9} and Table~\ref{tab:transitions} summarize the recurrent destinations.
	
	\begin{table*}[t]
		\centering
		\small
		\caption{Top recurrent uncataloged destinations from each cataloged state under noise, with change in vortex number and harmonic order.}
		\label{tab:transitions}
		\begin{tabular}{lccccc}
			\toprule
			Source $A_i$ & Destination $A_j$ & $P(A_j\,|\,A_i,\text{noise})$ & $\Delta N$ & $\Delta m^*$ & Direction \\
			\midrule
			S0 $(+,1,0)$   & $(+,3,1)$ & 0.30 & $+2$ & $+1$ & core gain \\
			S1 $(-,1,0)$   & $(-,3,1)$ & 0.33 & $+2$ & $+1$ & core gain \\
			S2 $(-,3,2)$   & $(-,3,1)$ & 0.72 & $0$  & $-1$ & harmonic decay \\
			S3 $(-,4,3)$   & $(-,3,1)$ & 0.30 & $-1$ & $-2$ & core loss + harmonic decay \\
			S4 $(+,5,2)$   & $(+,5,1)$ & 0.43 & $0$  & $-1$ & harmonic decay \\
			S5 $(-,7,3)$   & $(-,7,1)$ & 0.33 & $0$  & $-2$ & harmonic decay \\
			\bottomrule
		\end{tabular}
	\end{table*}
	
\begin{figure*}[t]
		\centering
		\includegraphics[width=\textwidth]{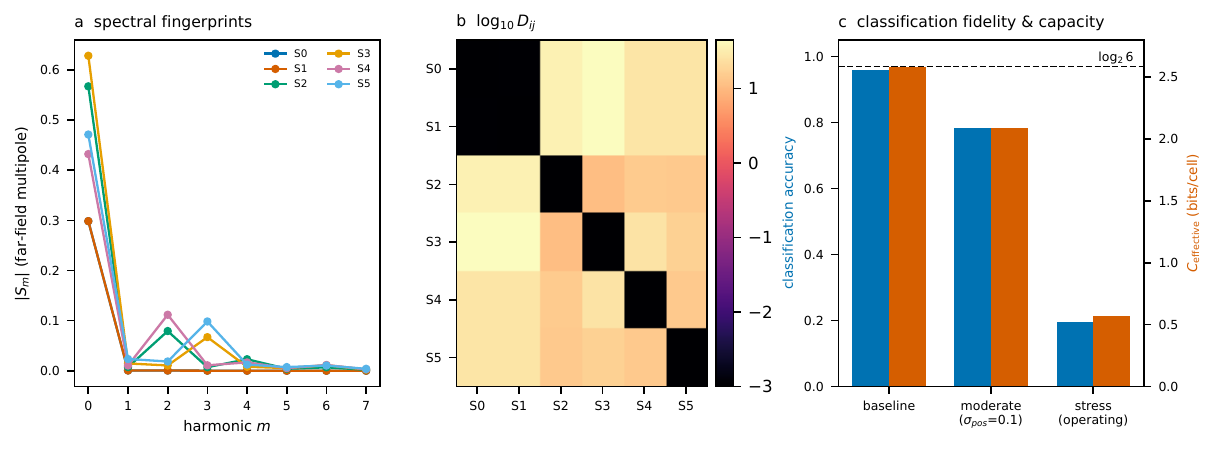}
		\caption{Non-destructive spectral readout and information capacity. \textbf{a}, Mean far-field multipole spectrum $|S_m|$. \textbf{b}, $\log_{10}$ pairwise spectral separation $D_{ij}$; chirality-degenerate pair excluded. \textbf{c}, Classification accuracy and effective capacity $\Ceff$ at the three operating points.}
		\label{fig:fig4}
	\end{figure*}

	Non-destructive readout uses the far-field circulation-multipole spectrum. Figure~\ref{fig:fig4}a shows the mean spectra; Fig.~\ref{fig:fig4}b shows pairwise spectral separation $D_{ij}$. All pairs except the chirality-degenerate monopoles (S0, S1) exceed $D\approx11$. The chirality bit is read independently via the sign of $\Gamma$. Figure~\ref{fig:fig4}c shows classification accuracy and effective information capacity.
	
	Six states correspond to $C_{\rm ideal}=\log_2 6=2.585$ bits/cell. Under noise, the achievable capacity $\Ceff$ (Blahut–Arimoto optimization) is $2.585$ bits/cell at baseline, $2.14$ bits/cell at moderate noise, and $0.54$ bits/cell under combined stress. We also report throughput metrics including write-verify-retry rates (see Supplementary Information).
	
	\subsection{Stability and basin geometry}
	\label{sec:stability}
	
	We computed linear stability in two formulations. First, the reduced positional subsystem (fixed cluster circulations and fixed hold current): every codeword is linearly stable, with $\max\Reall\,\lambda$ ranging from $-0.888$ (S0, S1) to $-5.30$ (S2). Second, the full coupled position-circulation Jacobian (circulation allowed to respond dynamically): the two monopoles remain stable, but all four multi-vortex codewords acquire a weak positive-real-part mode (Table~\ref{tab:instability}). However, the instability time $\tau_{\rm inst}$ exceeds the hold duration by over an order of magnitude, so the weak instability does not cause operational failure on the timescale of a single cycle.
	
	\begin{table*}[t]
		\centering
		\small
		\caption{Full coupled instability rate $\lambda_+$, instability time $\tau_{\rm inst}$, and operational hold duration. All four multi-vortex states are operationally robust despite $\lambda_+>0$; the ordering does not rank with basin radius.}
		\label{tab:instability}
		\begin{tabular}{lccccl}
			\toprule
			State & $\lambda_+$ & $\tau_{\rm inst}$ & $t_{\rm hold}$ & $\tau_{\rm inst}/t_{\rm hold}$ & Operational outcome \\
			\midrule
			S2 & $+0.076$  & $13.2$ & $0.6$ & $22$  & robust: $r_{50}=6.67\,\sigma_c$ \\
			S3 & $+0.0044$ & $229$  & $0.6$ & $382$ & robust: $r_{50}=8.00\,\sigma_c$ \\
			S4 & $+0.153$  & $6.5$  & $0.6$ & $11$  & robust: $r_{50}=7.80\,\sigma_c$ \\
			S5 & $+0.123$  & $8.1$  & $0.6$ & $14$  & robust: $r_{50}=7.38\,\sigma_c$ \\
			\bottomrule
		\end{tabular}
	\end{table*}
	
\begin{figure*}[t]
		\centering
		\includegraphics[width=\textwidth]{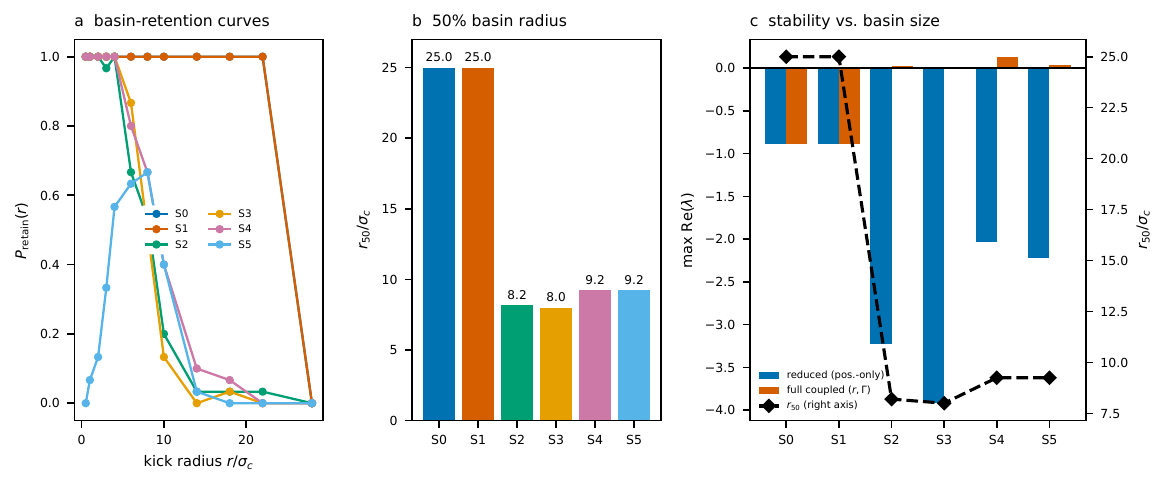}
		\caption{Basin-of-attraction robustness versus linear stability. \textbf{a}, Monte Carlo basin-retention curves $P_{\rm retain}(r)$. \textbf{b}, 50\%-basin radius $r_{50}$ (bootstrap intervals). \textbf{c}, Reduced and full coupled maximum eigenvalues (bars) alongside $r_{50}$ (diamonds): basin radius does not rank with either eigenvalue set.}
		\label{fig:fig7}
	\end{figure*}

	Basin-of-attraction robustness was measured directly via Monte Carlo: we applied a single instantaneous kick of radius $r$ to the cluster centroids and computed the fraction of trials that return to the target signature, $P_{\rm retain}(r)$. We extract the 50\%-basin radius $r_{50}$ (Fig.~\ref{fig:fig7}a,b). The two monopoles have $r_{50}\approx25\,\sigma_c$; the multi-vortex states have $r_{50}$ between $6.67$ and $8.00\,\sigma_c$ (Table~\ref{tab:basin}). The $r_{50}$ ordering differs from both the reduced and full coupled stability rankings (Fig.~\ref{fig:fig7}c, Table~\ref{tab:basin}): S2 has the strongest reduced stability but the smallest basin; S4 has the weakest coupled stability but a larger basin than S2. The basin radius therefore contains finite-amplitude information that is absent from either local stability spectrum.
	
	\begin{table*}[t]
		\centering
		\small
		\caption{Basin radius $r_{50}$ versus linear stability (reduced and full coupled). Rank among multi-vortex states (1 = most stable / largest basin) shows inversion: S2 ranks 1 in reduced stability but 4 in $r_{50}$; S4 ranks 4 in coupled stability but 2 in $r_{50}$.}
		\label{tab:basin}
		\begin{tabular}{lcccccc}
			\toprule
			State & $r_{50}/\sigma_c$ & $\max\Reall\,\lambda$ (reduced) & $\max\Reall\,\lambda$ (full) & Rank (red.) & Rank (full) & Rank ($r_{50}$) \\
			\midrule
			S0 & $\approx25.0$ & $-0.888$ & $-0.888$ & n/a & n/a & n/a \\
			S1 & $\approx25.0$ & $-0.888$ & $-0.888$ & n/a & n/a & n/a \\
			S2 & $6.67^{+1.62}_{-1.00}$ & $-5.30$ & $+0.076$ & 1 & 2 & 4 \\
			S3 & $8.00^{+0.67}_{-0.73}$ & $-3.92$ & $+0.0044$ & 3 & 1 & 1 \\
			S4 & $7.80^{+0.82}_{-0.68}$ & $-3.58$ & $+0.153$ & 4 & 4 & 2 \\
			S5 & $7.38^{+0.63}_{-0.38}$ & $-4.27$ & $+0.123$ & 2 & 3 & 3 \\
			\bottomrule
		\end{tabular}
	\end{table*}
	
	We also tested whether basin size follows from the nearest saddle point in the reduced system. We located index-1 saddles and computed a proxy basin radius; neither the proxy nor the saddle distance correlated with true $r_{50}$ (Pearson $r=+0.03$ and $-0.05$, $n=4$). This mismatch is consistent with the fact that the dominant escape channels for three of the four multi-vortex states involve circulation decay, which is absent in the fixed-circulation reduced system. Thus basin geometry is a property of the full coupled dynamics.
	
	\subsection{Retention and endurance}
	\label{sec:retention}
	
	We measured retention under active hold and passive decay. Under active hold, circulation is stable to small fluctuations (relative fluctuation 0.24–5.0\% over 2.5 time units). Once hold current is removed, $|\Gamma(t)|$ decays exponentially with an essentially state-independent e-folding time $\tau_{\rm passive}\approx0.475$ (dimensionless). This device is actively stabilized, not passively non-volatile -- retention requires continuous power.
	
	We define an operationally robust codeword as one satisfying $\Psucc(\sigma_{\rm pos}/\sigma_c=0.1)>0.5$ and $P_{\rm survival}(100)>0.9$. All six cataloged codewords pass this test (Table~\ref{tab:operational}). We tested endurance over 100 consecutive write–hold–read–erase cycles for all six codewords, repeated over 10 independent seeds (60 trajectories total). Every replicate completed all 100 cycles with zero failures for every codeword, giving a pooled per-cycle survival estimate $P_{\rm survival}=1.00$ (95\% Wilson interval $[0.996,1.00]$). Within the tested 100-cycle protocol, no systematic endurance failure was observed, although substantially longer cycling remains to be examined.
	
	\begin{table*}[t]
		\centering
		\small
		\caption{Catalogued versus operationally robust states. All three independent criteria agree: every cataloged codeword passes.}
		\label{tab:operational}
		\begin{tabular}{lcccc}
			\toprule
			State & $\Psucc(\sigma_{\rm pos}/\sigma_c{=}0.1)$ & $r_{50}/\sigma_c$ & $P_{\rm survival}(100)$ & Operationally robust \\
			\midrule
			S0 & 0.74 & $\approx25.0$ & 1.00 & yes \\
			S1 & 0.80 & $\approx25.0$ & 1.00 & yes \\
			S2 & 0.70 & 6.67 & 1.00 & yes \\
			S3 & 1.00 & 8.00 & 1.00 & yes \\
			S4 & 0.72 & 7.80 & 1.00 & yes \\
			S5 & 0.82 & 7.38 & 1.00 & yes \\
			\bottomrule
		\end{tabular}
	\end{table*}
	
	\section{Discussion}
	\label{sec:discussion}
	
	The stability and basin calculations reveal a clear separation between infinitesimal and finite-amplitude robustness. Basin radius $r_{50}$ does not rank with $\max\Reall\,\lambda$ in either formulation; thus basin geometry is an independent observable that must be measured directly. This holds even when upgrading from a linear spectrum to a full nonlinear saddle-point search, as long as circulation is held fixed. The failure of the saddle-point proxy is mechanistically informative: the dominant escape channels for three of four multi-vortex codewords involve circulation decay, which is excluded from the reduced system.
	
	The results also separate four operational requirements: reachability, reproducibility, robustness, and endurance. All six cataloged codewords satisfy all four in this study, but the hierarchy remains a useful general framework. The readout is non-destructive and uses boundary harmonic spectra; chirality provides an independent signed channel. The device is actively refreshed and requires continuous hold power; passive retention remains an open challenge.
	
	From an engineering perspective, the device geometry separates fabrication-fixed parameters (well count, positions, sign pattern) from electrically written ones (rim current, hold duration). In the proposed architecture, the multilevel state space is set by the fixed magnetic geometry and accessed electrically through the rim-current protocol. Candidate material platforms include high-mobility two-dimensional electron systems (graphene, GaAs) or liquid metals. A full energy calibration requires fixing a current scale and resistance (see Supplementary Information). The error mode is predominantly detectable erasure, favoring a write-verify architecture.
	
	\section{Conclusions}
	\label{sec:conclusions}
	
	We have shown that a self-organized vortex configuration in a boundary-driven electron fluid supports six reproducible codewords (2.585 bits/cell at baseline) with quantified noise robustness and endurance. Local eigenvalue spectra therefore do not, by themselves, specify the finite perturbation tolerance of the vortex-memory states considered here. The device requires active hold power; passive retention remains an open experimental challenge. Further validation requires materials-specific energy calibration, continuum-PDE comparison, and longer endurance tests.
	
	\section*{Supplementary Material}
	
	See the Supplementary Information for the full continuum-to-reduced derivation chain, extended-data tables (all tested noise conditions, kick radii, stability eigenvalues, uncataloged outcomes, saddle-point analysis), and a detailed comparison with existing memory technologies.
	
	\begin{acknowledgments}
		The authors thank the developers of the open-source scientific Python ecosystem (NumPy, SciPy, Matplotlib). No external funding was received.
	\end{acknowledgments}
	
	\section*{Author Declarations}
	
	\subsection*{Conflict of Interest}
	The authors declare no competing interests.
	
	\subsection*{Author Contributions}
	\textbf{M. Rabiu}: Conceptualization; Formal analysis; Investigation; Methodology; Software; Validation; Visualization; \textbf{S. S. Abukari}: Writing--original draft; \textbf{M. Amekpewu}: Writing--review \& editing.
	
	\section*{Data Availability Statement}
	
	The simulation code and data files supporting this study are available from the corresponding author upon reasonable request.
	
	\bibliographystyle{apsrev4-2}

\begin{thebibliography}{99}
		
		\bibitem{ye2021overview} Ye, P.~D. \emph{et al.} Overview and outlook of emerging non-volatile memories. \emph{MRS Bull.} \textbf{46}, 946–958 (2021).
		
		\bibitem{golod2015single} Golod, T., Iovan, A. \& Krasnov, V.~M. Single Abrikosov vortices as quantized information bits. \emph{Nat. Commun.} \textbf{6}, 8628 (2015).
		
		\bibitem{xiong2022ferroelectric} Xiong, H., Li, Y., Zhang, S. \& Wang, J. Ferroelectric vortex memory device concept. \emph{Front. Phys.} \textbf{9}, 791019 (2022).
		
		\bibitem{goto2021twin} Goto, K., Nakajima, K. \& Notsu, H. Twin vortex computer in fluid flow. \emph{New J. Phys.} \textbf{23}, 063051 (2021).
		
		\bibitem{bozkaya2011numerical} Bozkaya, N. \& Tezer-Sezgin, M. Numerical simulation of magnetohydrodynamic flow in a square cavity. \emph{Int. J. Numer. Methods Fluids} \textbf{67}, 1264–1282 (2011).
		
		\bibitem{bandurin2016negative} Bandurin, D.~A. \emph{et al.} Negative local resistance caused by viscous electron backflow in graphene. \emph{Science} \textbf{351}, 1055–1058 (2016).
		
		\bibitem{krishnakumar2017superballistic} Krishna Kumar, R. \emph{et al.} Superballistic flow of viscous electron fluid through graphene constrictions. \emph{Nat. Phys.} \textbf{13}, 1182–1185 (2017).
		
		\bibitem{newton2001nvortex} Newton, P.~K. \emph{The N-Vortex Problem: Analytical Techniques} (Springer, 2001).
		
		\bibitem{moller2020few} Möller, M., Gaida, J.~H., Schäfer, S. \emph{et al.} Few-nm tracking of current-driven magnetic vortex orbits using ultrafast Lorentz microscopy. \emph{Commun. Phys.} \textbf{3}, 36 (2020).
		
		\bibitem{blahut1972} Blahut, R. Computation of channel capacity and rate-distortion functions. \emph{IEEE Trans. Inf. Theory} \textbf{18}, 460–473 (1972).
		
		\bibitem{arimoto1972} Arimoto, S. An algorithm for computing the capacity of arbitrary discrete memoryless channels. \emph{IEEE Trans. Inf. Theory} \textbf{18}, 14–20 (1972).
		
		\bibitem{guckenheimer1983} Guckenheimer, J. \& Holmes, P. \emph{Nonlinear Oscillations, Dynamical Systems, and Bifurcations of Vector Fields} (Springer, 1983).
		
		\bibitem{strogatz1994} Strogatz, S.~H. \emph{Nonlinear Dynamics and Chaos: With Applications to Physics, Biology, Chemistry, and Engineering} (Perseus Books, 1994).
		
		\bibitem{bhati2016dram} Bhati, I., Chishti, Z., Lu, S.-L. \& Jacob, B. DRAM refresh mechanisms, penalties, and trade-offs. \emph{IEEE Trans. Comput.} \textbf{65}, 108–121 (2016).
		
		\bibitem{morley2020mhd} Morley, N.~B. \& Roberts, P.~H. MHD in liquid metals: from astrophysics to industry. \emph{Magnetohydrodynamics} \textbf{56}, 123–138 (2020).
		
		\bibitem{imec2023sotmram} imec. Imec's extremely scaled SOT-MRAM devices show record-low switching energy and virtually unlimited endurance. Press release (Dec. 13, 2023).
		
	\end{thebibliography}

	\appendix
	
	\section{Simulation methods}
	\label{app:methods}
	
	The Supplementary Information gives the full continuum-to-reduced derivation chain. Here we summarize the point-vortex dynamics and protocols.
	
	\subsection{Point-vortex dynamics}
	
	Point vortices $i=1,\ldots,N(t)$ at position $\bm r_i$ with circulation $\Gamma_i$ evolve under:
	\begin{widetext}
		\begin{equation}
			\dot{\bm r}_i = \bm v_i^{\rm (BS)} + \bm v_i^{\rm (img)} + \Omega_I(t)\,\hat{\bm z}\times\bm r_i \;-\; \mu_{\rm pos}(I)\,\bm r_i \;-\; \frac{1}{\Gamma_i}\nabla U_{\rm pin}(\bm r_i)\times\hat{\bm z},
		\end{equation}
	\end{widetext}
	where $\bm v_i^{\rm (BS)}$ is the regularized Biot–Savart velocity, $\bm v_i^{\rm (img)}$ is the Milne–Thomson image contribution, $\Omega_I(t)=\kappa_{\rm rot}I(t)$ is the rim-current solid-body rotation, $\mu_{\rm pos}(I)=\kappa_{\mu,\rm pos}(\kappa_e I)^2$ is a drag, and $U_{\rm pin}(\bm r)=\kappa_{\rm pin}B_{\rm mag}(\bm r)^2$ is the magnetic pinning potential from the patterned wells. Circulation evolves as $\dot\Gamma_i = -\mu_{\rm loc}(\bm r_i,t)\Gamma_i + \alpha I_{\rm source}(t)\Gamma_i/\Gamma_{\rm tot}$, with local damping $\mu_{\rm loc}(\bm r,t)=\mu_{\rm visc}+\kappa_\mu[\kappa_e I(t)+B_{\rm mag}(\bm r)]^2$. Integration is explicit RK4 at $dt=2\times10^{-3}$.
	
	Write ($t\in[0,0.25]$): jittered 16-vortex rim ring, pinning off. Split ($t\in[0.25,0.40]$): one satellite vortex per well, pinning on. Hold ($t\in[0.40,0.40+t_{\rm hold}]$): active hold with $I_{\rm hold}$ or passive decay. Erase: sinusoidal current for 0.30 time units.
	
	Boundary-harmonic readout uses $S_m = |\sum_k \Gamma_k r_k^m e^{-im\theta_k}|$, $m=0,\ldots,7$.
	
	\subsection{Protocols}
	
	Discovery: 186 grid points, seed 0; confirmation: seeds 1–4, $\ge4/5$ hits. Robustness: $t_{\rm hold}=0.6$, $n=50$ trials/condition, noise applied at every RK4 step. Confusion matrix: $n=50$ trials at three operating points. Basin radius: Monte Carlo kicks at 12 radii, $n=30$ trials/radius, $r_{50}$ by interpolation; bootstrap 5000 replicates for uncertainty. Cycling: 100 consecutive cycles, repeated over 10 independent seeds.
	
	\subsection{Stability and saddle-point analysis}
	
	Reduced positional Jacobian: finite-difference of position-update velocity at fixed circulations and fixed hold current. Full coupled Jacobian: finite-difference of full $(\dot{\bm r},\dot{\bm\Gamma})$ right-hand side. Saddle-point search: Newton polishing of reduced flow to fixed points, bisection to find critical radii, classification by Jacobian sign structure.
	
	\section{Supplementary tables (excerpt)}
	
	Detailed tables of all noise conditions, basin radii, stability eigenvalues, transition networks, and comparison with existing memory technologies are provided in the Supplementary Information.
	
\end{document}